\documentclass[aps,prb,onecolumn]{revtex4}

\usepackage{epsfig}

\newcommand{\be}{\begin{equation}}
\newcommand{\ee}{\end{equation}}

\begin{document}

\newcommand{\GaMnAs}{${\rm Ga}_{1-x}{\rm Mn}_x{\rm As}$}

\title{Theory of Manganese-Manganese interaction in \mbox{Ga$_{1-x}$Mn$_x$As}}

\author{P. Redli\'{n}ski$^1$, G. Zar\'and$^2$ and B. Jank\'o$^1$}

\address{
$^1$ Department of Physics, University of Notre Dame, Notre Dame, Indiana 46556\\
$^2$ Institute of Physics, Budapest University of Technology and
Economics, H-1521 Budapest, Hungary }

\begin{abstract}
We investigate the interaction of two Mn ions in the dilute magnetic
semiconductor  \GaMnAs{}  using the variational  envelope wave
function approach within the framework of six band model of the
valence band. We find that the effective interaction between the Mn
core spins at a typical separation $d$  is strongly anisotropic for
active Mn concentrations less than $x_{\rm active} = 1.3$~\%, but it
is almost isotropic for shorter distances ($d< 13$~\AA). As a
result, in unannealed and strongly compensated samples strong
frustration effects must be present.  We also verify that an
effective Hamiltonian description can be used in the dilute limit,
$x_{\rm active} < 1.3$~\%, and extract the parameters of this
effective Hamiltonian.
\end{abstract}


\maketitle

\section{Introduction}
Emergence of semiconductor with ferromagnetic properties\cite{Dietl,
Ohno} in (III-Mn)V materials leads to possibility of integration of
processing and magnetic storage in a single device. \GaMnAs{} is
promising material for such devices although today Curie temperature
up to 140~K is not high enough\cite{Ku}. GaMnAs is an example of
(III-Mn)V family of materials which evolved from II-VI based diluted
magnetic semiconductors\cite{Furdyna1}. These materials were
extensively studied in the 80's. Especially from our point of
interest, spin orbit interaction was studied in
Ref.~\onlinecite{Lee} (see also references therein). Marriage of
ferromagnetic and electronic degree of freedom is important today
because manipulation of the spin in solid state physics starts to be
applied at room temperature (recently room temperature
ferromagnetism has been observed in GaMnN exhibiting Curie
temperature up to 370K reported in Ref.~\onlinecite{Reed}) with the
perspectives to build spin diode, spin transistor and eventually
quantum computers.

Ferromagnetic semiconductor like \GaMnAs{} is characterized by
presence of localized magnetic moments and simultaneously by
presence of mobile holes which mediate ferromagnetic indirect
exchange interaction between Mn moments\cite{Sanvito}. It was shown
that substitutional Mn atom (Mn substitute Ga) forms localized
magnetic moment of 5/2 which come from its five 3d-electrons.
Additionally, divalent Mn atom substituting trivalent Ga atom
introduces one hole and forms effective mass acceptor. This
idealistic scenarios is not realized in real live and not all
intentionally incorporated Mn atoms come into Ga
sites\cite{Goldman}. \GaMnAs{} is highly compensated presumably due
to interstitials\cite{Blinowski} and antisite defects (As
substituted for Ga). It means that one to one correspondence between
Mn moments and number of holes is broken; For typical Manganese
concentration $x$=5\% number of holes $p$=0.3 per Mn atom is
expected\cite{Yu}.

There are two natural limits for which simulations are performed:
heavily\cite{Brey, Fiete2, Timm} and weakly doped
regimes\cite{Durst, Berciu2}. In the case of heavy doping one
considers valence-band holes moving in disordered potential of
defects. In this limit Zar\'{a}nd and Jank\'{o}\cite{Zarand}
suggested that frustration effects may be important for the magnetic
properties of the \GaMnAs{}. Using spherical approximation of the
valence band and performing mostly analytical calculations they
showed that spin anisotropy can suppress magnetization. Calculations
made by Brey and G\'{o}mez-Santos\cite{Brey} using full 6-band
valence-band Hamiltonian\cite{Abolfath} show that spin anisotropy is
smaller that predicted in Ref.~\onlinecite{Zarand}. In the second
limit of light doping the local impurity levels overlap only weakly
forming impurity bands in the gap. In this case tight binding model
is used\cite{Berciu2, Wang} or local density approximation approach
is applied\cite{Sanvito}. This limit is also interesting from
metal-insulator transition point of view\cite{Berciu, Timm2, Yu}.

There is some controversy in the literature. For example the
calculations of Ref.~\onlinecite{Fiete2,Brey,Berciu}
and~\onlinecite{Konig} are only valid in the high concentration
limit of the holes. Furthermore, most previous calculations ignored
the very large Coulomb potential of the negatively charged Mn
impurity, which actually provides the largest energy scale in the
problem\cite{Timm,acceptorGaAs}. This Coulomb potential can be
treated non-perturbatively in the very dilute limit, where an
impurity band picture applies
\cite{Berciu2,Fiete,Fiete3,Timm2,Wang}. However, the spin-orbit
coupling being large, one has to incorporate this also in realistic
calculations. While the spin-orbit coupling has been completely
neglected in Ref.~\onlinecite{Berciu2}, Ref.~\onlinecite{Fiete} took
into account the effects of spin-orbit coupling only within the
framework of the so-called spherical approximation\cite{BL}, where
the spin-orbit splitting between the spin $j=3/2$ and spin $j=1/2$
valence bands is taken to infinity, and therefore anisotropy effects
are overestimated.

In this paper we shall investigate the interaction between two Mn
ions, using the {\em full six-band
Hamiltonian}\cite{Lutt,Dietl2,Abolfath}, which is thought to give a
good approximation for the valence band excitations over a wide
energy range. To determine the effective interaction, we shall
compute  the spectrum of 'molecular orbitals' on the two Mn ions
through the application of variational  methods. These molecular
orbitals of the MnMn dimers are close analogues of the molecular
orbitals of the  $H_2^{+}$ molecule\cite{bookH2+}. As we shall see,
the energy of these molecular orbitals depends on the separation of
the two Mn ions and the direction of their spin, which we treat as
classical variables. Knowledge of the spin-dependence of  this
molecular spectrum allows us to  compute the effective interaction
between the Mn spins provided that there is a single bound hole on
the ${\rm Mn}_2$ 'molecule'.

As we shall also see,  both the approaches of
Ref.~\onlinecite{Berciu2} and Ref.~\onlinecite{Fiete} fail: While at
large separations the structure of the molecular orbitals and the
effective interaction is roughly captured by the spherical
approximation of Ref.~\onlinecite{Fiete2} (giving very large
anisotropies), for small Mn-Mn separations spin-orbit coupling
effects are not substantial, and the Mn-Mn interaction is
essentially isotropic, though the degeneracy of the bound hole
states is different from the one obtained through the naive spin 1/2
approach of Ref.~\onlinecite{Berciu2}. We shall also check if it is
possible to quantitatively describe the spectrum of the  ${\rm
Mn}_2$ molecule by an effective interaction, as proposed in
Ref.~\onlinecite{Fiete}. A detailed analysis of the spectrum reveals
that such an effective model description only works for large Mn-Mn
separations, $d>13$~\AA, corresponding to an active Mn
concentration, $x_{\rm active} < 1.3$~\%.

The paper is structured as follows. We will first consider
Mn-dimer\cite{Wang} problem using envelope wave function approach.
This approach is justified by the fact that when a small portion of
Gallium atoms is substituted randomly by manganese atoms we expect
that the band model of GaAs will be applicable to \GaMnAs{}, too.
After calculating energy states of the dimer we will map them into
the effective spin Hamiltonian\cite{Zarand, Zhou} and discuss
results of this mapping. We also will discuss spin anisotropy
effects.

\section{Variational Calculations}

To investigate the electronic structure of an ${\rm Mn}_2$
'molecule', we write the Hamiltonian describing the interaction of a
hole with the dimer as
\begin{equation}
{H}= H_{KL} - \sum_{i=1,2}\Bigl(\,\frac{e^2}{\epsilon r_i} +
V_c(r_i)\Bigr)
+ \sum_i J(r_i) \vec S_i\cdot {\vec s}_h \;. \label{HamAcceptor}
\end{equation}
Here the Kohn-Luttinger Hamiltonian $H_{KL}$ describes the kinetic
energy of the holes in the valence band within the envelope function
approach\cite{Lutt,Fishman}, and its detailed form is given in
Appendix~\ref{app:lutt}. The Coulomb interaction with the negatively
charged Mn cores is accounted for by the terms ${e^2}/{\epsilon
r_i}$, with $\epsilon=10.67$ denoting the dielectric constant of
GaAs host semiconductor, and $r_i = |\vec r-\vec R_i|$  being the
distance between the two Manganese ions at positions $\vec R_i$
($i=1,2$) and the hole at position $\vec r$. The so-called central
cell corrections\cite{acceptorGaAs}, $V_c(r_i)$, are used to take
into account the atomic potential in the vicinity of the Mn core
ions. For this central cell correction we used the simple Gaussian
form, $V_c(r) = V_c\;e^{-r^2/r_{c}^2}$. Finally, the last term in
Eq.(\ref{HamAcceptor}) takes into account the exchange interaction
with the two  Mn core spins, with   $\vec S_1$ and  $\vec S_2$ being
 spin \mbox{5/2} operators corresponding to half-filled $d$-levels.
These core spins shall be replaced by  classical spins in our
calculations, and we shall only treat the hole-spin
quantum-mechanically.

To regularize the  the  exchange interaction in
Eq.~(\ref{HamAcceptor}) we shall use a slightly  smeared delta
function,
\begin{equation}
 J(r) = \frac{J_{pd}}{(2\pi r_{pd}^2)^{3/2}}
\,e^{-{r}^2/2r_{pd}^2}\;, \label{eq:exchange}
\end{equation}
where $r_{pd}$ is a cut-off in the range of the inter-atomic
distance. While a microscopic derivation results in  a slightly more
complicated form for exchange interaction\cite{Timm,Fiete2},
Eq.~(\ref{eq:exchange}) can be also used as long as the spatial
scale of the exchange interaction, $r_{pd}$, is small
enough\cite{Fiete2,Brey}.

To determine the spectrum of Eq.~(\ref{HamAcceptor}) we used a
variational approach, where we expressed the wave function of the
holes in the following form,
\begin{eqnarray}
|\Psi\rangle &=&
\sum_{\mu=1}^6 \sum_{i,j,k}\,C_{ijk}^{\mu}\,\phi_i(\alpha x)\phi_j(\alpha y)\phi_k(\alpha z)|\mu\rangle, \\
\phi_i(\alpha
x)&=&\sqrt{\frac{\alpha}{\sqrt{\pi}2^ii!}}\,e^{-\alpha^2x^2/2}H_i(\alpha
x).\label{r12}
\end{eqnarray}
Here the $H_i$'s denote Hermit polynomials, and the label $\mu$
refers to the six spinor components (see Appendix~\ref{app:lutt}).
The minimization was  performed over the parameter $\alpha$ as well
as over the linear coefficients, $C_{ijk}^{\mu}$,
($i,j,k=0...N_{max}-1$). In our calculations, we used a cut-off
$N_{\rm max}=8$, but we also  checked that increasing the number of
basis  states did not change substantially our results for the Mn-Mn
separations discussed. The big advantage of using Hermitian
polynomials as basis states is that for these states one is able to
evaluate all matrix elements of the Hamiltonian analytically.

So far we did not discuss the value of the parameters $V_{c}$,
$r_{c}$, $J_{pd}$, and $r_{pd}$, characterizing the the central cell
correction and exchange interaction. While these phenomenological
parameters are not fully determined, they are not completely
arbitrary either: they must be chosen in such a way that the model
reproduces the experimentally observed binding energy of a hole,
$E_{Mn} \approx 111\; {\rm
meV}$\cite{beGaMnAs1,beGaMnAs2,acceptorGaAs}, and the spin splitting
between the bound hole states generated by the core Mn spin, $E_{\rm
split}\approx 12\; {\rm meV}$\cite{beGaMnAs2,Fiete}. Therefore, we
first performed calculations for a single Mn ion, and finally chose
a combination of parameters, \mbox{$J_{pd}=3.7\;{\rm eV}$~\AA$^3$},
\mbox{$r_{pd}=0.125$~\AA}, \mbox{$V_c=2668\;{\rm meV}$} and
\mbox{$r_c=2.23$~\AA}, which reproduced the above low energies
correctly.

\begin{figure}[t]
\centering
\includegraphics[width=9cm]{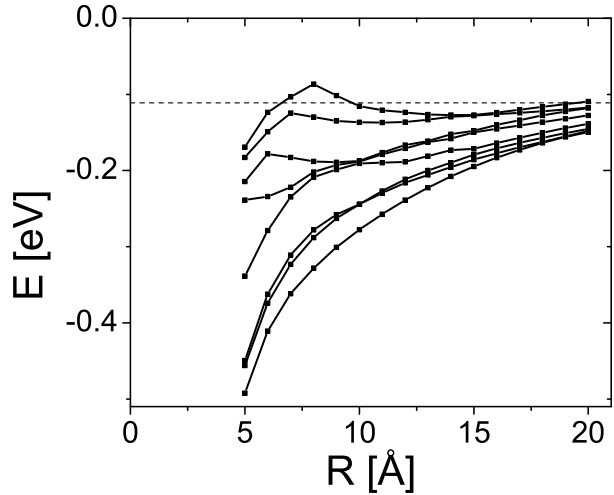}
\caption{Eight lowest lying energy states of the hole localized on
the two Mn ions as a function of the distance $R$ between the ions.
Both Mn spins  are pointing in the $z$-direction. The binding energy
of the Mn acceptor (-0.111~eV) is indicated by the dashed line.
Below $R=10$~\AA~the spectrum becomes  more complicated then for
$R>10$~\AA. \label{8LowestStates}}
\end{figure}

Having these parameters in hand,  we returned to the problem of an
${\rm Mn}_2$ dimer and computed the  spectrum of the 'molecule' as a
function of the separation $R$ between  the two Mn impurities and
their spin orientations. Our calculations were performed for $R$ in
the range 5~\AA< $R$ < 20~\AA. Note that for  GaAs lattice constant
$a$ is approximately $a=5.6$~\AA and the nearest separation between
substitutional atoms in face centered cubic lattice of about 4~\AA.
For the sake of simplicity, we performed computations only for the
simplest case where the positions of the two Mn ions were aligned
along the $z$-direction. In general, however, the spectrum and the
effective interaction of the two Mn ions depends on the orientation
of the Mn bond as well as the orientation of the spins.

Fig.~\ref{8LowestStates} shows the evolution of the eight
lowest-lying states of the ${\rm Mn}_2$ dimer for the case when both
spins point in the $z$-direction. In the absence of the Mn core
spin, the ground state of an isolated Mn ion would be fourfold
degenerate\cite{acceptorGaAs,BL}. This degeneracy is slightly split
by a presence of the spin: A classically treated  Mn core spin lifts
this fourfold degeneracy, and results in a small splitting of these
four states, which are still well-separated from the rest of the
spectrum of the ion.

If we now take approach  two Mn ions to each other, then these four
states of the two ions hybridize, and give rise to eight (bonding
and anti-bonding) molecular orbitals. For small enough
concentrations
it is enough to keep only these eight states to build an effective
impurity band Hamiltonian for \GaMnAs{}\cite{Fiete}. These eight
states are well-separated from the rest of the spectrum for $R>
8$~\AA, but at around $R\approx 8$~\AA~level crossings occur, and
the whole envelop function approximation breaks dow for separations
smaller than this.

\begin{figure}[ht]
\centering
\includegraphics[width=9cm]{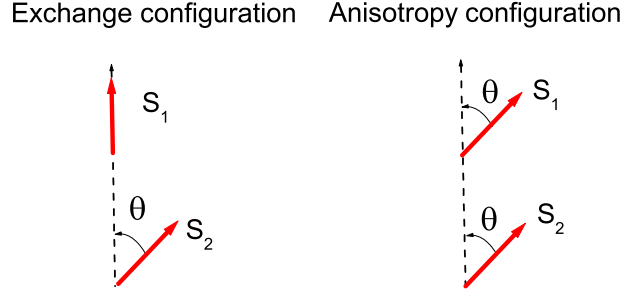}
\caption{Configurations used to determine the anisotropy energy and
the exchange energy. In the 'exchange configuration' one Mn-spin
($\vec{S}_1$) is pointing in the $z$-direction and the second
Manganese spin ($\vec{S}_2$) is rotated into the $-z$-direction. In
the 'anisotropy configuration' both spins $\vec{S}_1$ and
$\vec{S}_2$ are parallel and rotated by the same angle $\theta$. In
our calculations spins were rotating in the YZ plane.
\label{Configurations}}
\end{figure}

We can also determine the spectrum of the dimer as a function of the
orientation of the two Mn spins, $\vec{S}_1$ and  $\vec{S}_2$, and
their separation $R$. This spin orientation-dependent spectrum is
very useful to estimate the exchange energy and the anisotropy
energy between the two Mn ions. In particular, the spin
orientation-dependence of the energy of the lowest-energy bonding
state just gives the effective interaction energy of the two spins
provided that there is a single hole on the dimer (roughly
corresponding to a hole fraction $f=0.5$~\%).

\begin{figure}[ht]
\begin{center}
\includegraphics[width=9cm,clip]{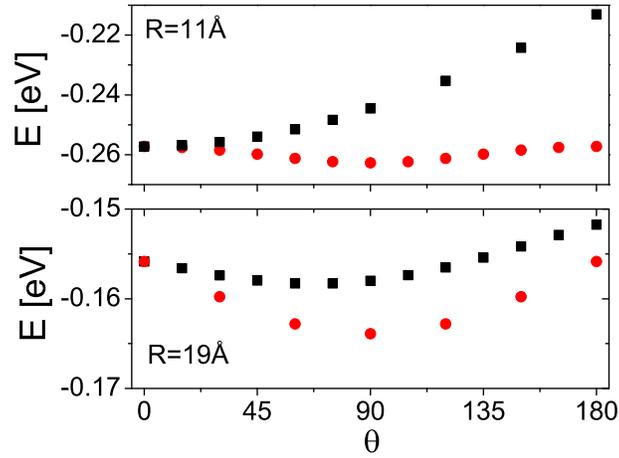}
\end{center}
\caption{Dependence of the binding  energy of a hole for two
separations ($R=11$~\AA~and $R=19$~\AA) as a function of angle
$\theta$. Circles  stand for  anisotropy configuration, squares
belong to the 'exchange configuration'. For large separations the
anisotropy energy is comparable or larger than the exchange energy,
while for small separations the exchange energy dominates.
Anisotropy always prefers an Mn spin orientation perpendicular to
the bond, $\theta=90$. \label{figR11R19}}
\end{figure}

We determined the spectrum of the molecule for two different
('exchange' and 'anisotropy') types of spin configurations shown in
Fig.~\ref{Configurations}. The corresponding binding energies of a
single hole are shown in Fig.~\ref{figR11R19}. For a separation
$R\sim 10$~\AA, the anisotropy energy is very large (in the range of
$\sim 100\; {\rm K}$) but is only about  $20$~\% of the exchange
energy, which tends to align the two spins ferromagnetically,
$\theta=0^{\circ}$. The anisotropy energy prefers an orientation
perpendicular to the bond, $\theta=90^{\circ}$, in agreement with
the RKKY results of Ref.~\onlinecite{Zarand}, but in disagreement
with the results of Fiete {\em et al.}\cite{Fiete}, where an 'easy
axis' anisotropy has been found. (The origin of this disagreement
between the results of Ref.~\onlinecite{Fiete} and the present
computations is unclear). For larger separations the anisotropy
energy does not decrease too much, but  the exchange energy drops by
about a factor of 10 for separations $\sim 20$~\AA. As a result, the
anisotropy plays a crucial role for Mn-Mn separations $R>\;13$~\AA.

\begin{figure}[ht]
\includegraphics[width=9cm]{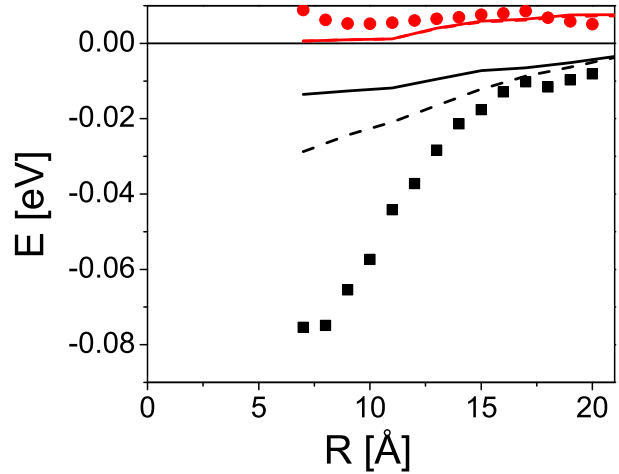}
\caption{Exchange energy ($E_{\rm exch}$, squares) and non-collinear
anisotropy energy ($E_{\rm anis}$, circles) as a function of
distance $R$ between the two Mn ions. Anisotropy energy is positive
and exchange energy is negative indicating that Mn spins tend to
line up ferromagnetically  in the plane perpendicular to the
direction joining them. Results obtained using the effective model
are also plotted as lines. Full lines corresponds to $G$=4~meV in
the effective model  and dashed lines to $G$=10~meV.  Anisotropy
turns out to be the same for both choices of $G$ so the second
dashed line is not seen. The effective model give a reasonable
description of the interaction only for separations $R>16$~\AA.
\label{figExchAnisVsR}}
\end{figure}

Similar to Refs.~\onlinecite{Brey, Fiete2}, we can define the
exchange and anisotropy energies as
\begin{equation}
E_{\rm exch} \equiv E_{\uparrow\uparrow} -
E_{\downarrow\downarrow}\;,\phantom{nn} E_{\rm anis} \equiv
E_{\uparrow\uparrow} - E_{\rightarrow\rightarrow}\;,
\end{equation}
with the arrows indicating the spin direction with respect to the
$z$-axis. These energies are plotted in Fig.~\ref{figExchAnisVsR}.
Note that for separations $R > 13$~\AA~(corresponding to an active
Mn concentration $x_{\rm active} < 2.2$~\%)  the anisotropy energy
becomes comparable to the exchange energy. In this range
orientational frustration effects discussed in
Refs.~\onlinecite{Zarand} and \onlinecite{Fiete} are expected to be
important.

\section{Effective Hamiltonian}
\label{sec:Heff}

As we already mentioned in the previous Section, in the very dilute
limit the eight lowest lying states of a pair of Mn ions can be
constructed from the four lowest-energy acceptor states of the Mn
ions, since these are well separated from the rest of the
spectrum\cite{Fiete}. Furthermore, the exchange coupling being small
compared to the separation between the above-mentioned four states
and higher-energy excitations, one may attempt to replace it by a
local interaction, which we write in the second quantized form as
\be J_{pd}\dots  \to \sum_{i=1,2} G\; \vec S_i \cdot \vec F_i =
 \sum_{i=1,2} \sum_{\mu,\nu}G\; \vec S_i\cdot (c^\dagger_{i\mu} \vec F_{\mu\nu}
c_{i\nu}) \;, \label{eq:G} \ee where $\vec F_i$ denotes the
effective spin  of the bound hole on ion $i$, and $G\approx 5 \;
{\rm meV}$  is an effective coupling that has been determined from
 infrared spectroscopy\cite{beGaMnAs2}.
The matrices $ \vec F_{\mu\nu}$ above are just  spin 3/2 matrices
spanning the four-dimensional Hilbert space of the lowest-lying
acceptor states of each Mn ion, and  the operators
$c^\dagger_{i\mu}$ create a hole on these four states at ion $i=$
with spin component $F^z = \mu$.

Let us set the exchange coupling $J_{pd}$ to zero for a moment.
Then, in the spirit of tight binding approximation, the most general
Hamiltonian can be written within the subspace of these eight
acceptor states as
\begin{eqnarray}
H^{\rm eff}_{{\rm Mn}-{\rm Mn}}& = &\sum_{\mu,\nu=1\dots 4}
t_{\mu\nu}(\vec R) \left( c^\dagger_{1,\mu}c_{2,\nu} + {\rm h.c.}
\right ) \label{eq:H_eff}
\\
\nonumber &+& \sum_{\scriptstyle i=1,2 \atop \scriptstyle\mu,\nu}
 c^\dagger_{i,\mu} \; \left( K_{\mu\nu}(\vec R) + \delta_{\mu\nu}E(\vec R) \right)  \;c_{i,\nu}
\;.
\end{eqnarray}
Note that the hopping  terms $t_{\mu\nu}$ are not diagonal in $F_z$
as a consequence of spin-orbit coupling. Furthermore, spin-orbit
coupling also generates anisotropy terms, $K_{\mu\nu}$ which,
however, turn out to be relatively small and can usually be
neglected. All parameters in Eq.~(\ref{eq:H_eff}) depend on the
relative position $\vec R$ of the two Mn ions.

\begin{figure}[ht]
\includegraphics[width=9cm]{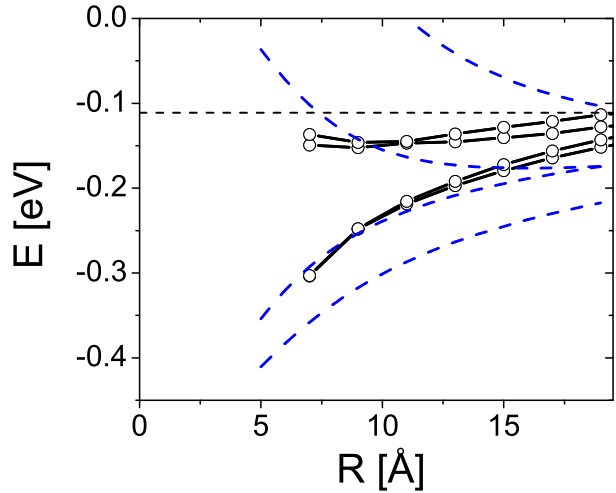}
\caption{Lowest lying energy states of a manganese pair (lines with
symbols) as a function of distance between manganese atoms $R$ in
the absence of the core spins, e.g., $J_{pd}=0$. Dashed lines
indicate the results of Ref.~\onlinecite{Fiete}. Each line is
two-fold degenerate. \label{figWithOutJpd}}
\end{figure}

Within the spherical approximation used in Ref.~\onlinecite{Fiete},
the angular dependence of $t_{\mu\nu}$ and $K_{\mu\nu}$ is trivially
given by spin 3/2 rotation matrices, and the effective Hamiltonian
 Eq.~(\ref{eq:H_eff}) becomes  just a function of four parameters
that depend only on the distance $R$ between the two Mn ions.  In
the six band model, on the other hand, the  numerous parameters of
the effective Hamiltonian are typically complicated functions of
$\vec R$. However, if the positions of the two Mn ions are aligned
along the $z$-direction, than  the effective Hamiltonian simplifies
a lot due to time reversal symmetry and the $C_4$ symmetry of the
Hamiltonian (remember that we set $J_{pd} \equiv 0$ for the time
being), and the effective Hamiltonian simplifies as
\begin{eqnarray}
H_{Mn-Mn}^{eff}&=&\sum_{\nu=1\dots
4}t_{\nu}(R)(c^{\dag}_{1,\nu}c_{2,\nu}+h.c.)
\\
\nonumber
&+&\sum_{i,\nu}[K(R)(\nu^2-\frac{5}{4})+E(R)]c^{\dag}_{i,\nu}c_{i,\nu}\;.
\end{eqnarray}
To determine the  four parameters, $t_{1/2}$, $t_{3/2}$, $K$ and
$E$, we performed a variational calculation with the previously
obtained parameters $V_c$, $r_c$, and $r_{pd}$,
 but setting $J_{pd} \equiv 0$. The results are
shown in Fig.~\ref{figWithOutJpd}, where for comparison, we also
show the results obtained within the spherical approximation of
Ref.~\onlinecite{Fiete}.

While the two approximations give a qualitatively similar spectrum
at large separations, the spherical approximation badly fails at
short separations. There within the six-band model calculation we
obtain a bonding and an anti-bonding state, each of which has an
approximate fourfold spin degeneracy. In other words, at short
separations the spin-orbit coupling is not very important. In
contrast, the spherical approximation  gives a large spin-orbit
splitting of the states even for small separations.

This difference is easy to understand: for small Mn-Mn separations,
the bound states are composed from valence band states of {\em all
energies}. However, while the spherical approximation gives a
reasonable approximation for the band structure in the close
vicinity of the $\Gamma$ point, it badly fails at these high
energies, where it overestimates the effect of spin-orbit coupling.
At large separations, on the other hand, the effective Hamiltonian
is determined by the {\em tails} of the wave functions, which are,
in turn, composed from small momentum valence band excitations. The
structure of these states at large distances is therefore roughly
captured by spherical approximation, though the results
quantitatively differ. These observations parallel the ones of
Ref.~\onlinecite{Fiete2}, where we have shown using the RKKY
approximation that spin-orbit coupling becomes important only for
large Mn-Mn separations within that approach too.

\begin{figure}[ht]
\includegraphics[width=9cm]{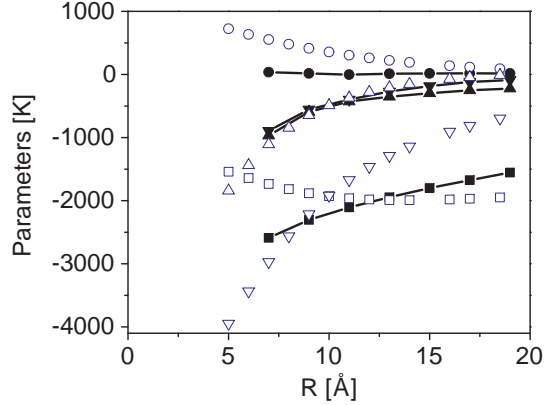}
\caption{Four parameters $E$ (squares), $K$ (circles), $t_{12}$
(down triangle) and $t_{32}$ (up triangle) in two approximations:
calculated by variational approach with constrains $J_{pd}=0$ (lines
with symbols) and taken from Ref.~\onlinecite{Fiete} (only symbols).
Energy is given in Kelvin: 100~K $\approx$ 8.8~meV.\label{compEKT}}
\end{figure}

The extracted effective parameters $E$, $K$, $t_{12}$ and $t_{32}$
are shown in Fig.~\ref{compEKT}. The largest parameter is $E(R)$,
which asymptotically approaches the single ion binding energy.  The
approach to this asymptotic value is dominated by the Coulomb
interaction, and $E(R)$ falls off as \be E(R) \approx E(\infty) -
\frac{e^2}{\epsilon R}\;. \ee

The most important effect of spin-orbit coupling is that $t_{1/2}$
and $t_{3/2}$ become very different for large separations, and
$|t_{1/2}| > |t_{3/2}|$ implying that holes with spin aligned
perpendicular to the bond  can hop more easily. This is the ultimate
reason why spin-orbit coupling favors an Mn spin-orientation {\em
perpendicular} to the bond. Note that this is very different from
the result obtained within the spherical approximation, where
$|t_{1/2}| < |t_{3/2}|$, and therefore the Mn spins are preferably
aligned along the bond. In fact, the relation $|t_{1/2}| >
|t_{3/2}|$ is also favored by intuition: the weight of the valence
band orbitals with $m=0$ angular momentum in the $d$-channel  is
larger for the $F_z = \pm 1/2$ states than for  $F_z = \pm 3/2$
states. Since the $m=0$ orbitals are the ones that point along the
Mn-Mn bond, one expects a larger hopping in the $\mu=\pm 1/2$
channels.

\begin{figure}[ht]
\includegraphics[width=9cm]{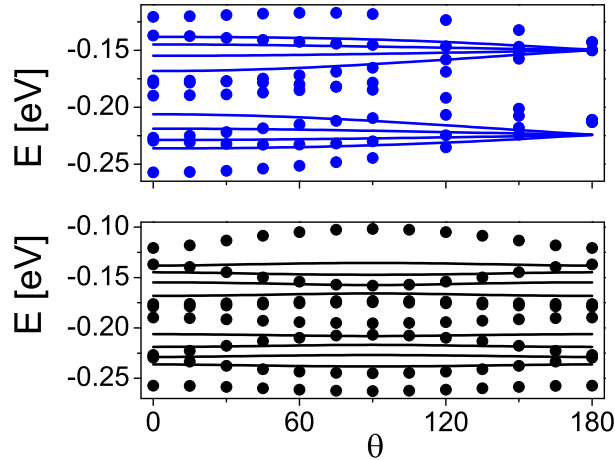}
\caption{Energies of the eight lowest lying states of the Mn dimer
vs. $\theta$ of a effective model (line) and full calculations
(points) at $R=11$~\AA. The upper panel corresponds to exchange
configuration while the lower panel shows results for the
'anisotropy configuration'. \label{ExchAnisR11}}
\end{figure}

\begin{figure}[ht]
\includegraphics[width=9cm]{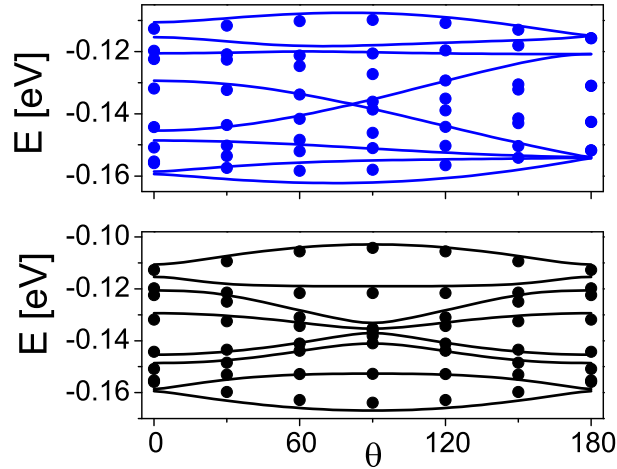}
\caption{Energies of the eight lowest lying states of the Mn dimer
vs. $\theta$ of a effective model (lines) and full calculations
(circles) at $R=19$~\AA. The upper panel corresponds to exchange
configuration while the lower panel shows results for the
'anisotropy configuration'. \label{ExchAnisR19}}
\end{figure}

We shall now proceed and test if the effective Hamiltonian defined
by Eqs.~(\ref{eq:G}) and (\ref{eq:H_eff}) gives indeed a reasonable
description of the Mn-Mn interaction also in the presence of the Mn
spin and for arbitrary  spin orientation. To this purpose we
computed the spectrum of the Mn dimer from the effective Hamiltonian
and compared it to the results of a full variational calculation
with $J_{pd}\ne0$. The results for separations $R=11$~\AA~and
$R=19$~\AA~are shown in Figs.~\ref{ExchAnisR11} and
\ref{ExchAnisR19}, respectively. Clearly, while for $R=11$~\AA~the
effective Hamiltonian is unable to capture the details of the full
spectrum, for $R=19$~\AA~it gives a very satisfactory description of
it.

To make a more quantitative comparison, we introduced the average
error of the effective Hamiltonian as
\begin{equation}\label{testchi}
\chi^2 \equiv\Bigl\langle \sum_{j=1}^{8} (E_{j} -E_{j}^{eff})^2
\Bigr\rangle_{\rm conf}\;,
\end{equation}
where the average is over all spin configurations and the eight
lowest-lying states. The quantity $\chi$ thus characterizes the
typical deviation from the real (variationally obtained) spectrum.
The obtained value of $\chi$ is shown in Table~\ref{fitEKT}. The
parameters ($E$, $K$, $t_{3/2}$, and $t_{1/2}$) have been extracted
from a variational calculation with $J_{pd} = 0$ in the way
discussed above. However, a small error  in the largest parameter
$E$ results in a large change in $\chi$, without changing the
internal structure and excitation spectrum of the dimer. (The
anisotropy or exchange energies, {\em e.g.}, are completely
independent of the value of $E$.) Therefore,  to eliminate this
error, we optimized the value of $E$ by slightly changing it,  $E
\to E_{\rm best}$. The optimal values of $E$, and the corresponding
$\chi$'s are shown in Table~\ref{fitEKT}. For $R\ge 15$~\AA~the
effective Hamiltonian gives a rather good description, and $\chi$ is
only about $\sim 10$~\% of the overall width of the spectrum, $\sim
2 |t_{1/2}|$. For $R = 11$~\AA, however, the relative error  goes
above $20$~\%, and the the effective model's spectrum hardly
resembles to the one obtained through a full variational calculation
with $J_{pd}\ne 0$ (see Fig.~\ref{ExchAnisR11}).

\begin{table}[ht]
\begin{tabular}{|c|c|c|c|c|c|}\hline
  \textit{R} [\AA]         & \textbf{11}   & \textbf{13}   & \textbf{15}   & \textbf{17}   & \textbf{19}   \\\hline
  \textit{E} [meV]         & -181 & -168 & -155 & -144 & -134   \\\hline
  \textit{K }[meV]         & -0.2 & -1.0 & -1.4 & -1.5 & -1.4   \\\hline
  $t_{3/2}$ [meV]           & -34.3& -23.3& -15.9& -10.2& -7.6    \\\hline
  $t_{1/2}$ [meV]           & -36.9& -30.4& -25.6& -21.6& -19.3   \\\hline
    $G$ [meV]               & 4.0 &   4.0   &  4.0    &  4.0    & 4.0 \\\hline
  $E_{\rm best}$ [meV]         & -187 & -172 & -159 & -147 & -136   \\\hline
 $\chi$ [meV]            & 15.0 & 8.0 & 5.8 & 4.1 & 3.9       \\\hline
\end{tabular}
\caption{ The values of the effective parameters, $E$, $K$,
$t_{3/2}$ and $t_{1/2}$ for different $R$, extracted from a
calculation with $J_{pd} = 0$, and the error $\chi$ of the effective
model. The typical error $\chi$  has been computed after changing
the shift $E$ to its optimum  value,  $E_{\rm best}$, and using
$G=4$~meV. \label{fitEKT}}
\end{table}

The quality of the effective model approximation can be somewhat
further improved  by not extracting the values of $E$, $K$,
$t_{3/2}$, and $t_{1/2}$ from a $J_{pd}=0$ variational calculation,
but instead, considering them and also  $G$ as fitting  parameters
to minimize $\chi$. However, even after a full optimization, it is
impossible to satisfactorily describe the $R=11$~\AA~excitation
spectrum of the molecule for any spin configuration, while the $R\ge
13$~\AA~spectra are not substantially improved.

\section{Conclusion}

In this paper, we studied the 'molecular orbitals' of  two
substitutional Mn ions as a function of their position and spin
orientation within the six-band model, using a variational approach.
Similar to the results of Ref.~\onlinecite{Fiete2}, we find that
spin-orbit coupling effects are important if the separation between
the two Mn ions are large ($R> 13$~\AA). In this regime spin-orbit
coupling  induces an anisotropy in the spin-spin interaction that
prefers to align the Mn spins perpendicular to the Mn-Mn dimer's
direction. For smaller separations, however, although large,
spin-orbit coupling induced anisotropy
 does not seem to be crucial compared to the
ferromagnetic exchange interaction.  These results are in
qualitative agreement with those of Refs.~\onlinecite{Fiete2, Timm}
but clearly disagree with those of Ref.~\onlinecite{Brey}, where a
very small anisotropy has been found. This difference  originates
from the different cut-off scheme used in  Ref.~\onlinecite{Brey},
which, as discussed in Refs.~\onlinecite{Fiete2, Timm} in detail,
suppresses back-scattering processes if the cut-off parameter is not
chosen carefully enough.

The transition to  anisotropy-dominated regions occurs at relatively
small active manganese concentrations, $x_{\rm active} \approx
2$~\%.  We have to emphasize though that interstitial Mn ions
presumably bind to substitutional Mn ions, and therefore the active
Mn concentration can be substantially reduced compared to  the
nominal concentration $x$ of Manganese\cite{Yu, Blinowski}. For an
unannealed $x=6$~\% sample, {\em e.g.}, with only one third of the
Mn ions going to interstitial positions, the active Mn concentration
can be reduced down to $x_{\rm active} = 2$~\%, in which regime the
anisotropy of the effective spin-spin interaction is already very
significant.

Upon annealing, however, interstitial Mn ions presumably diffuse out
of the sample, leading to a higher active Mn concentration.
Therefore, in annealed samples the effective interaction between the
Mn core spins should be  much more isotropic, and the spin of the
active Mn ions forms a  fully aligned ferromagnetic state. These
results are in agreement with the experimental data of
Ref.~\onlinecite{Ku}, where in some of the samples with small Curie
temperature, the remanent  magnetization could be {\em
substantially} increased by a relatively small magnetic field,
clearly hinting to a generically non-collinear magnetic state. We
have to mention though that other possible mechanisms have also been
suggested to cause  non-collinear
ferromagnetic states\cite{Schliemann, Korzhavyi}. 

We also tested if one can construct a simple effective impurity-band
Hamiltonian  in terms of spin 3/2 holes hopping between the Mn sites
to describe \GaMnAs{} in the dilute limit, as proposed in
Ref.~\onlinecite{Fiete}. We find that such a description only makes
sense if  Mn-Mn separations are large enough, $R>13$~\AA, and
becomes reliable only for active Mn concentrations below $x_{\rm
active} \approx 1.5 $~\%. We also determined the parameters of the
effective model by performing calculations within the framework of
the six band model, and found that while  earlier calculations done
within the spherical approximation are qualitatively correct for
large Mn separations, they quantitatively {\em fail}  to reproduce
the spectrum obtained within the six band model. An interesting
result of these six-band model calculations is that the effective
hopping parameters turn out to be considerably {\em smaller} than
the ones found within the spherical approximation. This implies that
the impurity band may survive to somewhat larger concentrations,
$x_{\rm active} \approx 2 - 3$~\%, as also suggested by
angle-resolved photoemission spectroscopy, scanning tunneling
microscopy and optical conductivity experiments.

In summary, we find that an effective model description in the
spirit of Ref.~\onlinecite{Fiete} is only possible if the active Mn
concentration is less than about $x_{\rm active}=1.5$~\%. We find
furthermore that spin-orbit coupling induced anisotropy is extremely
large for concentrations \mbox{$x_{\rm active}<2$~\%}, and it likely
leads to the appearance of frustrated non-collinear states.

\appendix
\widetext

\section{Six-band Hamiltonian}
\label{app:lutt} For the sake of completeness, we give in this
Appendix the effective six-band Hamiltonian used o describe the top
of the valence band in our computations. Within this approach the
valence band holes wave function is a six-component spinor, and the
so-called  Kohn-Luttinger Hamiltonian acts on these six-component
spinors,\cite{Lutt, Dietl2, Abolfath}
\begin{equation}
\label{HL}
    H_{KL}=
\pmatrix{
  H_{hh}     & -c & -b & 0 & b/\sqrt{2} & c\sqrt{2} \cr
  -c^{\star} & H_{lh} &0 & b & -b^{\star}\sqrt{3/2} & -d \cr
  -b^{\star} & 0 & {H_{lh}} & {-c} & d & -b\sqrt{3/2} \cr
  {0} & {b^{\star}} & {-c^{\star}} & {H_{hh}} & -c^{\star}\sqrt{2} & b^{\star}/\sqrt{2} \cr
  b^{\star}/\sqrt{2} & -b\sqrt{3/2} & d^{\star} & -c\sqrt{2} & H_{so} & 0 \cr
  c^{\star}\sqrt{2} & -d^{\star} & -b^{\star}\sqrt{3/2} & b/\sqrt{2} & 0 & H_{so}}\;.
\end{equation}

\narrowtext In this expression $b$, $c$, $d$, $H_{h}$, $H_{l}$, and
$H_{so}$ denote the following differential operators,
\begin{eqnarray}
H_{h/l} & = &
\frac{\hbar^2}{2m_0}\bigl[[-(\gamma_1\pm\gamma_2)\left(\frac{d^2}{dx^2}+\frac{d^2}{dy^2}\right)
\nonumber
\\
&-&(\gamma_1\mp2\gamma_2)\frac{d^2}{dz^2}\Bigr],\\
H_{so} & = &
\Delta_{so}-\frac{\hbar^2}{2m_0}\gamma_1\left(\frac{d^2}{dx^2}+\frac{d^2}{dy^2}+\frac{d^2}{dz^2}\right),\\
b & = & \frac{\hbar^2}{2m_0}2\sqrt{3}\gamma_3\left(-\frac{d}{dx}+i\frac{d}{dy}\right)\frac{d}{dz},\\
c & = & \frac{\hbar^2}{2m_0}\sqrt{3}\left[\gamma_2\left(-\frac{d^2}{dx^2}+\frac{d^2}{dy^2}\right)+2i\gamma_3\frac{d}{dx}\frac{d}{dy}\right],\\
d & =
&-\frac{\hbar^2}{2m_0}\sqrt{2}\gamma_2\left[-2\frac{d^2}{dz^2}+\frac{d^2}{dx^2}
+\frac{d^2}{dy^2}\right],
\end{eqnarray}
and the Kohn-Luttinger Hamiltonian, Eq.~(\ref{HL}), was written in
the basis
of the following 6 spin states, 
\begin{equation}
\label{basis}
\begin{array}{l}
|j=\frac{3}{2};m_j=+\frac{3}{2}>, \\
|j=\frac{3}{2};m_j=-\frac{1}{2}>,\\
|j=\frac{3}{2};m_j=+\frac{1}{2}>,\\
|j=\frac{3}{2};m_j=-\frac{3}{2}>,\\
|j=\frac{1}{2};m_j=+\frac{1}{2}> \;\;\text{and}\\
|j=\frac{1}{2};m_j=-\frac{1}{2}>.
\end{array}
\end{equation}
The first four components of the spinors describe the so-called
heavy- and light hole bands on the top of the valence band of GaAs.
These four bands become degenerate at the $\Gamma$ point and have
$j=3/2$ character.  The last two components of the spinors describe
the spin-orbit split bands. These two bands have a spin $j=1/2$
character at the $\Gamma$ point, where they are separated in GaAs by
an amount $\Delta_{\rm so} = 0.34\;{\rm eV}$ from the other four
bands.

The so-called Luttinger parameters, $\gamma_i$, must be determined
to give the best agreement with the experimentally observed band
structure of GaAs. In our calculations we used the set of parameters
$\gamma_1=7.65$, $\gamma_2=2.41$, and $\gamma_3=3.28$, frequently
used in the literature\cite{BL}.
%
%
In the so-called spherical approximation, where only the top four
bands are kept and the $\gamma_i$'s are set to $\gamma_1=7.65$, and
$\gamma_2=\gamma_3=2.93$\cite{BL}. In this approximation the
Hamiltonian acquires an SU(2) symmetry.

For the sake of completeness, let us give here the spin operators of
the holes which read in the above basis:
\begin{eqnarray}
    s_x=
\pmatrix{0 & 0 & \frac{1}{2\sqrt{3}}& 0 & \frac{1}{\sqrt{6}} & 0 \cr
        0 & 0 & \frac{1}{3} & \frac{1}{2\sqrt{3}} & -\frac{1}{3\sqrt{2}} & 0 \cr
        \frac{1}{2\sqrt{3}} & \frac{1}{3} & 0 & 0 & 0 & \frac{1}{3\sqrt{2}} \cr
        0 & \frac{1}{2\sqrt{3}} & 0 & 0 & 0 & -\frac{1}{\sqrt{6}} \cr
        \frac{1}{\sqrt{6}} & -\frac{1}{3\sqrt{2}} & 0 & 0 & 0 & -\frac{1}{6} \cr
        0 & 0 & \frac{1}{3\sqrt{2}} & -\frac{1}{\sqrt{6}} & -\frac{1}{6} &
        0},
\phantom{nnn}
   s_y=i
\pmatrix{0 & 0 & -\frac{1}{2\sqrt{3}}& 0 & -\frac{1}{\sqrt{6}} & 0
\cr
    0 & 0 & \frac{1}{3} & -\frac{1}{2\sqrt{3}} & -\frac{1}{3\sqrt{2}} & 0 \cr
        \frac{1}{2\sqrt{3}} & -\frac{1}{3} & 0 & 0 & 0 & -\frac{1}{3\sqrt{2}} \cr
        0 & \frac{1}{2\sqrt{3}} & 0 & 0 & 0 & -\frac{1}{\sqrt{6}} \cr
        \frac{1}{\sqrt{6}} & \frac{1}{3\sqrt{2}} & 0 & 0 & 0 & \frac{1}{6} \cr
        0 & 0 & \frac{1}{3\sqrt{2}} & \frac{1}{\sqrt{6}} & -\frac{1}{6} &
        0},
\nonumber
\end{eqnarray}

\begin{equation}
\label{sz}
    s_z=
\pmatrix{\frac{1}{2} & 0 & 0& 0 & 0 & 0 \cr
    0 & -\frac{1}{6} & 0 & 0 & 0 & -\frac{\sqrt{2}}{3} \cr
    0 & 0 & \frac{1}{6} & 0 & -\frac{\sqrt{2}}{3} & 0 \cr
    0 & 0 & 0 & -\frac{1}{2} & 0 & 0 \cr
    0 & 0 & -\frac{\sqrt{2}}{3} & 0 & -\frac{1}{6} & 0 \cr
    0 & -\frac{\sqrt{2}}{3} & 0 & 0 & 0 & \frac{1}{6}}.
\end{equation}


\end{document}